\documentclass[aps,prb,twocolumn,groupedaddress,showpacs]{revtex4}
\usepackage{graphicx}
\usepackage{natbib}

\begin{document}

\title{Emission of continuous coherent terahertz waves with tunable frequency by intrinsic Josephson junctions}

\author{Masashi Tachiki}
\affiliation{National Institute for Materials Science, 1-2-1 Sengen, Tsukuba 305-0047, Japan}

\author{ Mikio Iizuka, Kazuo Minami, Syogo Tejima, and Hisashi Nakamura }
\affiliation{ Research Organization for Information Science and Technology, 2-2-54 Nakameguro, Meguro-ku, Tokyo 153-0061, Japan }

\date{\today}

\begin{abstract}
We present a mechanism for emission of electromagnetic terahertz waves by simulation. High $T_{c}$ superconductors form naturally stacked Josephson junctions. When an external current and a magnetic field are applied to the sample, fluxon flow induces voltage. The voltage creates oscillating current through the Josephson effect and the current excites the Josephson plasma. The sample works as a cavity, and the input energy is stored in a form of standing wave of the Josephson plasma. A part of the energy is emitted as terahertz waves.
\end{abstract}


\maketitle

Continuous coherent terahertz waves have various applications in scientific field such as biology and information science. One of the hurdles for technological advancements in the terahertz region of electromagnetic wave is the development of sources for intense and continuous coherent terahertz waves. Therefore, we investigate a new mechanism for emitting intense continuous and frequency tunable terahertz waves.  In the high temperature superconductors, the strongly superconducting CuO$_{2}$ layers and insulating layers are alternatively stacked along the c-axis of the crystals and form a naturally multi-connected Josephson junction called intrinsic Josephson junction (IJJ). In the IJJ there appears a new excitation wave called Josephson plasma, the frequency of which is in the range of terahertz\cite{Tachiki1,Bulaevskii1}. The frequency appears in the region inside the superconducting energy gap and the Landau damping is very weak, and thus the excited plasma decays by emitting a terahertz electromagnetic wave.

For investigating an emission mechanism of terahertz electromagnetic wave from the IJJ, we use the following model shown by Figure 1. In Fig. 1 the IJJ is shown in green and the electrodes of a normal metal (for example gold) are shown in yellow.  An external magnetic field B applied in the direction of the y-axis induces fluxons in the direction.  The centers of fluxons are in the insulating layers. In this system, the superconducting and normal currents almost uniformly flow in the direction indicated by $\it J$ in Fig.1.  The fluxons flow in the direction of the x-axis with a velocity v and induce the flow voltage in the direction of the z-axis. These voltages creates the oscillating Josephson current along the z-axis by the Josephson effect, when temperature is low enough below $T_{c}$ and the superconducting current is smaller than the superconducting depairing current along the c-axis.  This oscillating current interacts strongly with the Josephson plasma due to the nonlinear nature of the system and intensively excites the Josephson plasma wave as shown later.  We use Bi$_{2}$Sr$_{2}$CaCu$_{2}$O$_{8+\delta}$ that is appropriate in the experiments, and apply a magnetic field and external currents around $J_{c}$ the critical current to the IJJ.  Then, the frequency of the plasma waves appears in the terahertz frequency range. The plasma wave is converted to an intense terahertz electromagnetic wave in the waveguide (dielectric) shown in orange in Fig. 1.

\begin{figure}
\includegraphics[width=16pc]{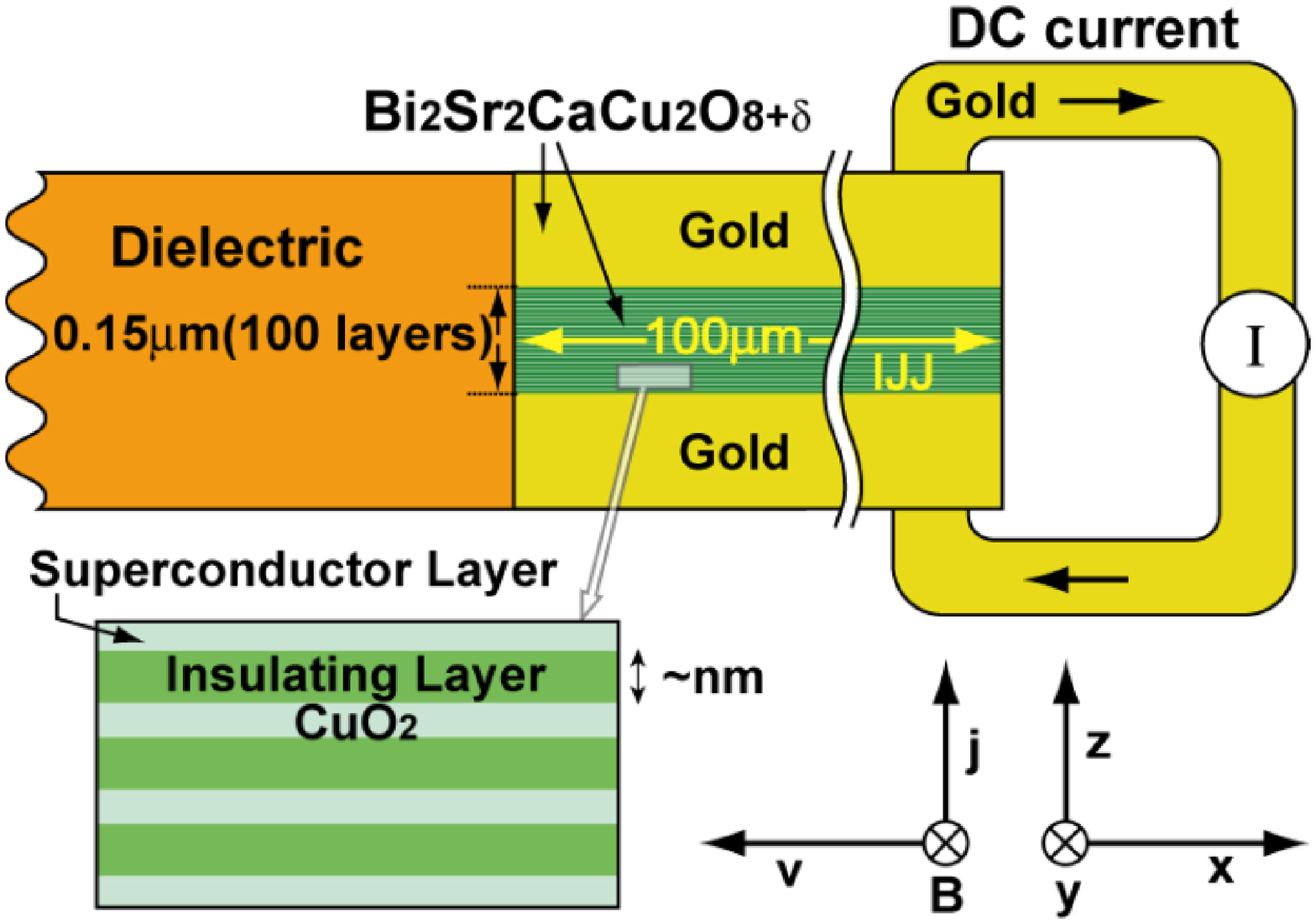}
\caption{ Schematic diagram of a prototype model for terahertz emission. 
Bi$_{2}$Sr$_{2}$CaCu$_{2}$O$_{8+\delta}$ forms the IJJ shown in green, and electrode gold plates form the top and bottom electrodes shown in yellow. A dielectric waveguide shown in orange extends from the left surface of the IJJ.}
\end{figure}

In accordance with the mechanism mentioned above, we now derive the equations for the simulation.  The superconducting order parameter of the $l$th layer is expressed as $\psi_{l}({\bf r},t)=\Delta_{l}e^{i\varphi_{l}({\bf r},t)}$.  Since superconductivity in the CuO$_{2}$ layers is strong at low temperatures, we assume $\Delta_{l}({\bf r},t)$ to be spatially and time independent.  In this system, the current along the z-axis perpendicular to the CuO$_{2}$ plane is given by a sum of the Josephson, quasi-particle, and displacement currents by

\begin{eqnarray}
J_{z,l+1,l}({\bf r},t)=J_{c}\sin \psi_{l+1,l}({\bf r},t)+\sigma_{c}E_{z,l+1,l}({\bf r},t)\nonumber\\
+ \frac{\varepsilon _{c}}{4\pi }\partial _{t}E_{z,l+1,l}({\bf r},t),
\end{eqnarray}
where $J_{c}$ is the critical current in zero magnetic field, $\sigma_{c}$ is the normal conductivity along the c-axis, and E$_{z,l+1,l}$ is the electric field between ($\it l$+1)th and $\it l$th layers along the z-axis. The $\psi_{l+1,l}$ is the gauge invariant phase difference defined as

\begin{equation}
\psi_{l+1,l}({\bf r},t) \nonumber\\
=\varphi_{l+1}({\bf r},t)-\varphi_{l}({\bf r},t)
-\frac{2\pi}{\phi_{0}} \int\limits_{z_{l}}^{z_{l+1}}{dzA_{z}} ({\bf r},z,t),
\end{equation}
with the vector potential $A_{z}({\bf r},z,t)$ and the flux unit $\phi_{0}$. For the superconducting current densities in the CuO$_{2}$ plane, we use the generalized London equations, since the Ginzburg-Landau parameter is very large in cuprate superconductors. We insert Eqs. (1) and (2) into Maxwell$^{\prime}$s equations taking account of the superconducting current in the $ab$-planes and the charging effect in the CuO$_{2}$ plane. Following the calculation procedure\cite{Tachiki2} and after cumbersome calculations, we have

\begin{eqnarray}
(1-\zeta\Delta^{(2)}) [(\partial_{t^{\prime}}^{2}\psi_{l+1,l}+\beta\partial_{t^{\prime}}\psi_{l+1,l}+\sin\psi_{l+1,l}) \nonumber\\
 +{\alpha}s^{\prime}(\partial_{t^{\prime}}(\rho_{l+1}^{\prime}-\rho_{l}^{\prime}) +\beta(\rho_{l+1}^{\prime}-\rho_{l}^{\prime}))]\nonumber\\
= \partial_{x^{\prime}}^{2} \psi_{l+1,l} + \partial _{y^{\prime}}^{2} \psi_{l+1,l},
\end{eqnarray}

\begin{equation}
s^{\prime}(1 - \alpha \Delta ^{(2)} )\rho_{l}^{\prime} = \partial t^{\prime}(\psi_{l + 1,l} - \psi _{l,l -1}).
\end{equation}
In the above equation, we use normalized units for length, time, and charge density in the $\it l$th CuO$_{2}$ plane, respectively defined as

\begin{equation}
x^{\prime} = \frac{x}{\lambda_{c}},  t^{\prime}=\omega_{p}t,  and \;  \rho_{l}^{\prime}= \frac{\lambda _{c}\omega_{p}\rho_{l}}{j_{c}},
\end{equation}
where $\rho_{l}$ is the charge density in the $\it l$th CuO$_{2}$ layer.  The plasma gap frequency is given by $\omega_{p}=c/(\sqrt{\varepsilon_c} \lambda_{c})$, $\varepsilon_{c}$ being the dielectric constant along the z-axis and $\lambda_{c}$ being the magnetic field penetration depth from the bc surface plane. The parameters in Eqs. (3) and (4) are defined as 

\begin{equation}
\zeta = \frac{\lambda_{ab}^{2}}{sd}, \alpha = \frac{\varepsilon_{c}\mu ^{2}}{sd}, s^{\prime} =\frac{s}{\lambda_{c}},  and \; \beta =\frac{4\pi \sigma_{c}\lambda_{c}} {\sqrt \varepsilon_{c} c} 
\end{equation}
where the $\it s$ and $\it d$ are the thickness of the superconducting and insulating layers, respectively. $\lambda_{ab}$ is the London penetration depth of the superconducting layer, and $\mu$ is the Debye screening length. The operator $\Delta^{(2)}$ is defined as $\Delta^{(2)} f_{l} = f_{l+1} - 2f_{l} + f_{l-1}$. We use Eqs. (3) and (4) in the IJJ and use Maxwell$^\prime$s equations in the dielectric to simulate the emission of terahertz wave.

If Eqs. (3) and (4) are linearized,  we obtain the Josephson plasma solution.  The Josephson plasma is a composite wave of the Josephson current and electromagnetic wave. The amplitudes of the Josephson current and the electric field are always parallel to the c-axis. Depending on the wave propagation directions parallel to the c-axis and the a-axis, there appear respectively the longitudinal and transverse plasma waves.  The transverse and longitudinal waves have been observed by Tamasaku et al.\cite{Tamasaku1}, and Matsuda et al.\cite{Matsuda1}, respectively.

Let us show how the plasma wave is converted to the electromagnetic wave at the interface between the IJJ and the dielectric.  As mentioned before, the plasma waves have the transverse and longitudinal components. However, the electromagnetic wave has only the transverse component.  Therefore, only the transverse plasma wave can convert into the electromagnetic wave at the interface.  When we solve Eqs. (3) and (4), we impose the following boundary condition.  To connect the Josephson plasma wave in the IJJ to the electromagnetic wave in the dielectric at the interface, we put the usual electromagnetic boundary condition; the electric and magnetic fields parallel to the interface are continuous at the interface. The electromagnetic wave in the dielectric is assumed to transmit freely to outer space at the end surface of the dielectric.  One possible way to achieve this condition is to make a gradation of the dielectric constant near the surface. It is assumed that the opposite surface of the IJJ is exposed to vacuum.

Keeping in mind that the IJJ is Bi$_{2}$Sr$_{2}$CaCu$_{2}$O$_{8+\delta}$ and the dielectric is MgO, we chose $\lambda_{ab}$=0.4$\mu$m, $\lambda_c$=200$\mu$m, s=3$\AA$, d=12$\AA$, $\mu$=0.6$\AA$, $\alpha$=0.1, $\beta$=0.01$\sim$0.05, and the number of layers=20$\sim$1000, and take the dielectric constants along the z-axis in the IJJ and that of the dielectric wave-guide to be $ \varepsilon$=10. We apply a magnetic field of 1Tesla along the y-axis.  We supply an external current from the gold electrodes shown in yellow in Fig. 1.  We change the normalized external current $J/J_{c}$ from 0.2 to 1.5 in step of 0.0125. The superconductivity in the top and bottom electrodes is assumed to penetrate to 0.075 $\mu$m.  We note that for the external currents mentioned above, the superconducting current part of $J$ is always less than $J_c$. The length of the IJJ is taken to be 100$\mu$m along the x-axis and the length of the dielectric is taken to be 50$\mu$m along the x-axis. For each external current, the time evolution is simulated until the system reaches a stationary state, that is normalized time t$^\prime$=600, and the emission power is calculated after that time.

We assume that the system is uniform along the y-axis and make two-dimensional calculation in the x-z plane.  We use the finite difference method to perform the numerical simulation. The simulation uses very large sized nonlinear equations heretofore difficult to compute; for a simulation using $10^{6}$ spatial cells in the x-z two-dimensional model, it would take two years to simulate one case of $10^{8}$ time-steps using a personal computer with a 2GHz processor. Therefore, we used the Earth Simulator that has a peak performance of 40 teraflops and we carried out this simulation in one day. We simulated a total of sixty cases. Therefore, in sixty days we carried out simulations that would have required one hundred and twenty years using a personal computer.

Since the electromagnetic wave is the transverse wave, only the node-less Josephson plasma wave along the z-axis (the transverse plasma) can be converted into the electromagnetic wave with the same frequency. The maximum stacking number that fulfill the condition is determined by the applied magnetic field, and it increases as the magnetic field and $\lambda_{ab}$ increase. We found that the stacking number that fulfill the condition is 100 layers under the magnetic field of 1 Tesla and $\lambda_{ab}$ of 0.4$\mu$m.

\begin{figure}
\includegraphics[width=20pc]{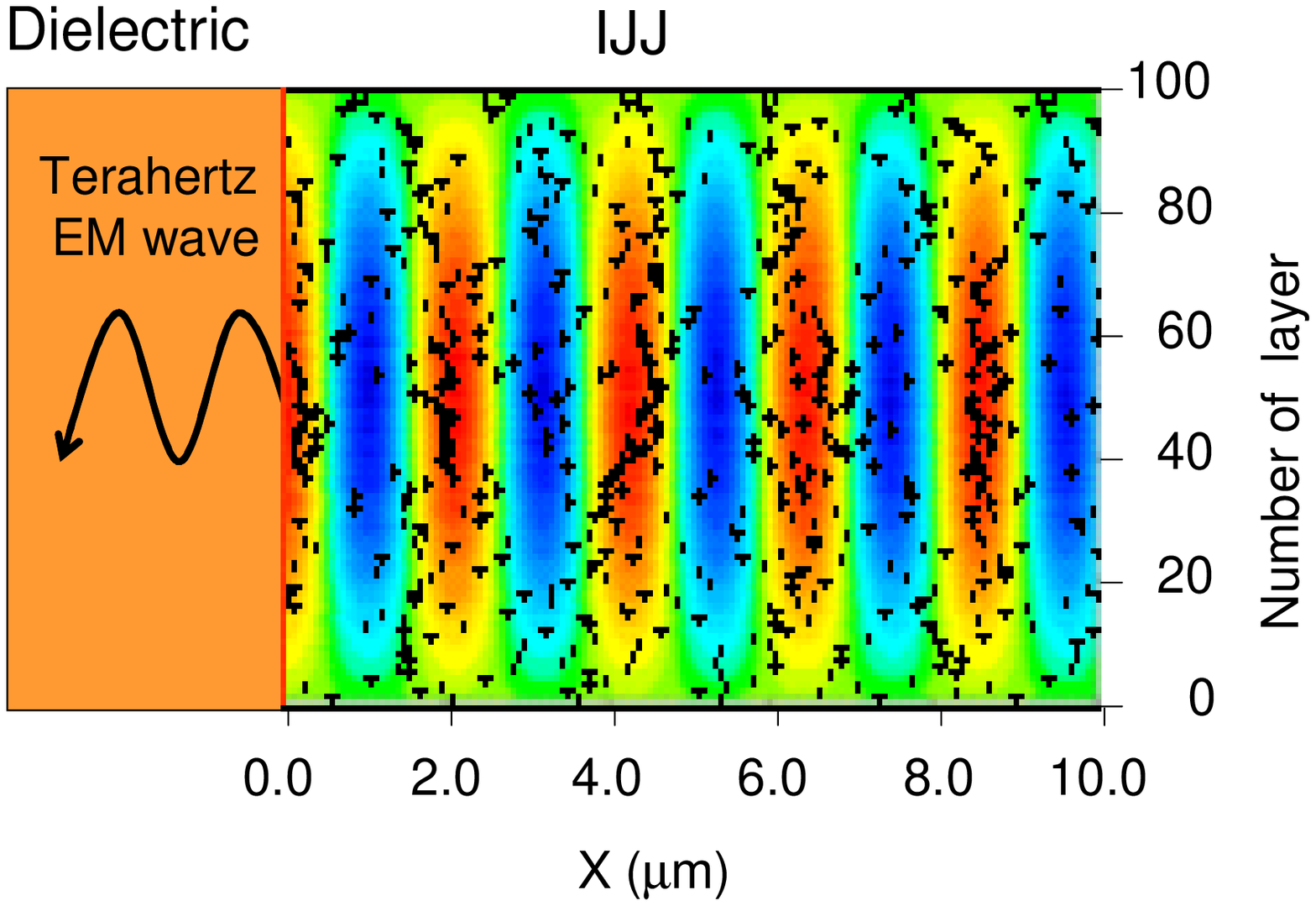}
\caption{A snapshot of moving fluxons and the oscillating electric fields of the standing plasma wave.}
\end{figure}

Figure 2 shows a snapshot of the moving fluxons and the oscillating electric field. In the figure the static and uniform electric field have been subtracted. The black circles show the centers of the fluxons. The fluxons are almost randomly distributed and move uniformly in the x-axis direction.  The speed of the fluxons is about 4$\%$ of the speed of light in the current $J/J_{c}$=0.6, $\beta$=0.02 and under the applied magnetic field of 1 Tesla. The magnetic field of a fluxon spreads in the range of $\lambda_{c}$ along the $ab$-plane. In the magnetic field of 1 Tesla, 120 fluxons are contained in the range. Therefore, the magnetic field of the moving fluxons are uniform and thus the fluxon flow create a spatially almost uniform voltage in the direction of the z-axis and the voltage induces the oscillating current due to the Josephson effect and the oscillating current excites the Josephson plasma. The plasma wave is reflected at the interface between the IJJ and the dielectric, and at the surface exposed to vacuum, and it forms a standing wave in the stationary state as shown in Fig. 2. In the figure red and yellow show the electric field directing to the z-axis direction, and blue shows the electric field directing to the -z axis direction.

\begin{figure}
\includegraphics[width=15pc]{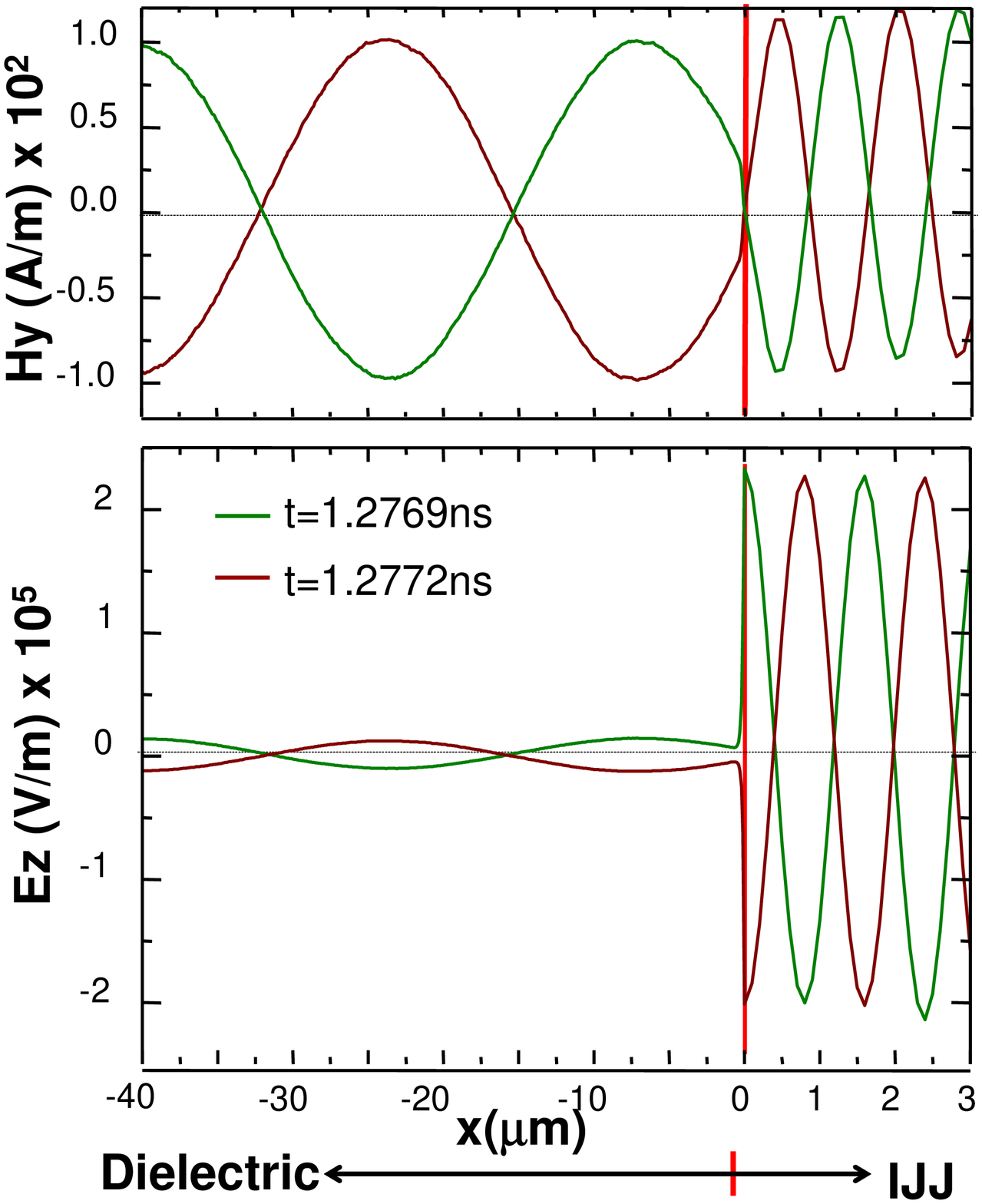}
\caption{A snapshot of the plasma wave that is a standing wave in the IJJ and the electromagnetic wave that is a propagating wave in the dielectric.}
\end{figure}

Figure 3 shows a snapshot of the electric and magnetic fields of the standing wave at the 50th insulating layer from the bottom surface in the direction z-axis. The standing wave of the electric field along the z-axis oscillates with 2.86 THz around a constant electric field for the normalized current current $J/J_{c}$=0.8 and for $\beta$=0.02. The red line shows the interface between the dielectric and the IJJ. The components of the electric and magnetic fields parallel to the interface fulfill the electromagnetic continuity condition. In Fig. 3 the green curves indicate the magnetic and electric fields at the time of 1.2769ns after the system attains to the stationary state and the brown curves indicate the magnetic and electric fields at the time of 1.2772ns. The belly of an electric field oscillation with the largest amplitude is pinned at the interface, inducing the large amplitude of the electric and magnetic field oscillations in the dielectric. The ratio of the electric field to the magnetic field is equal to that of the electromagnetic wave at 4$\mu$m from the interface. The emission mechanism of the coherent terahertz electromagnetic wave is similar to a laser mechanism. The IJJ itself works as a cavity.  The energy is stored in the form of a standing Josephson plasma in the IJJ.  A few percent of the energy is emitted as a terahertz electromagnetic wave from the IJJ to the dielectric through the interface. 

\begin{figure}
\includegraphics[width=16pc]{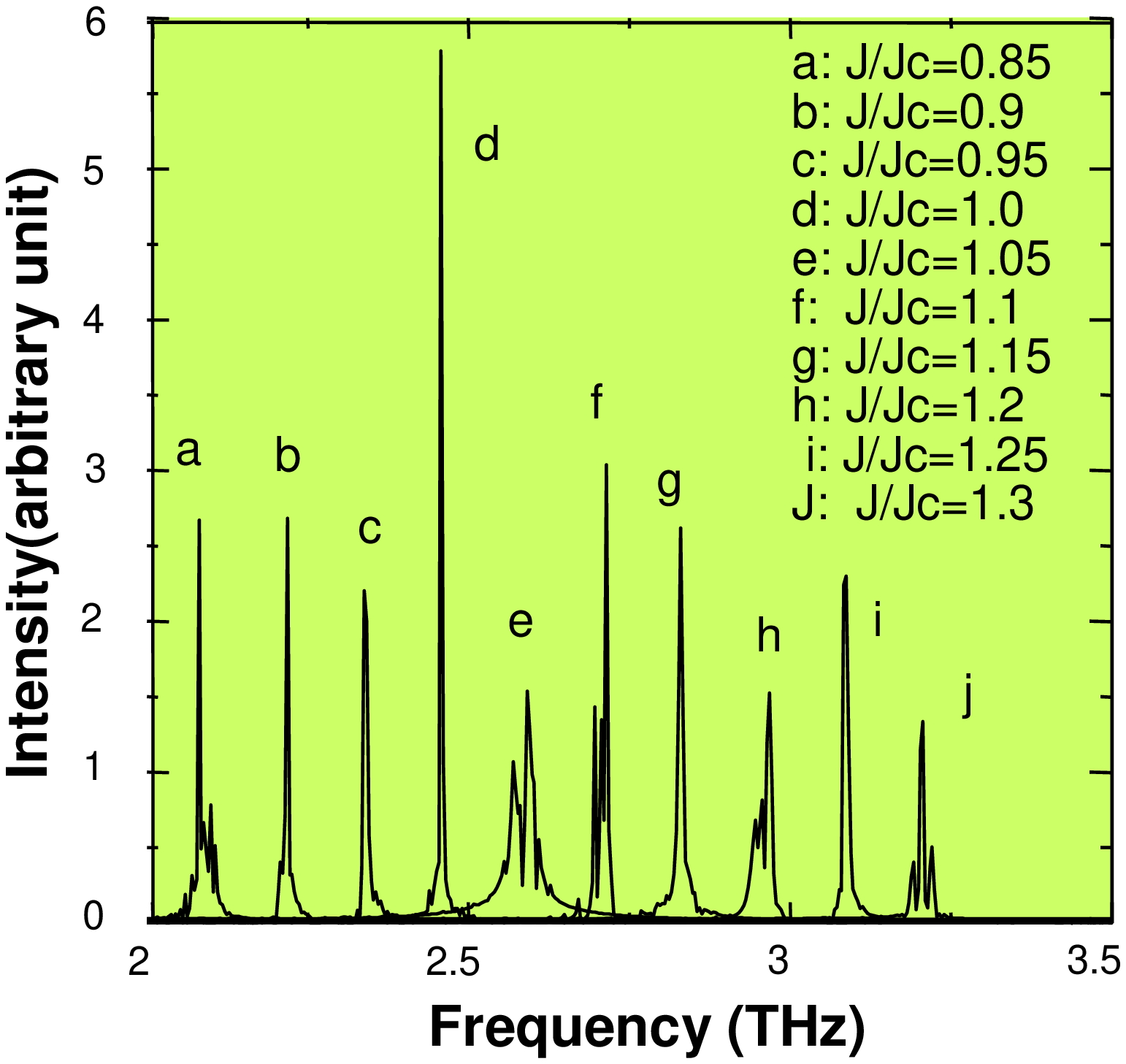}
\caption{The frequency spectra analyzed by FFT for 10 normalized currents. The 
intensities are those of electromagnetic waves in the dielectric. The intensity is obtained by
calculating the Poynting vector at a location of 4$\mu$m from the interface.}
\end{figure}

In Fig. 4 we show the frequency spectra calculated by FFT analysis of the electromagnetic wave at a location 4$\mu$m from the interface between the IJJ and the dielectric for the case of $\beta$=0.03. When we change the normalized current $J/J_{c}$ from 0.85 to 1.3, the frequencies of the electromagnetic waves change from 2.08 THz to 3.21 THz. The frequency spectra are very sharp. The frequency of the electromagnetic wave in the IJJ varies continuously by changing current $J/J_{c}$, indicating that emission of the sample is frequency tunable. Some peaks have satellites due to the frequency modulation of the excited plasma wave. The frequencies calculated by the simulation are equal to the AC Josephson frequency estimated by the fluxon flow voltage within the accuracy of 1.3$\%$.  This fact suggests that the plasma wave is excided by the oscillating current through the AC Josephson effect.

\begin{figure}
\includegraphics[width=16pc]{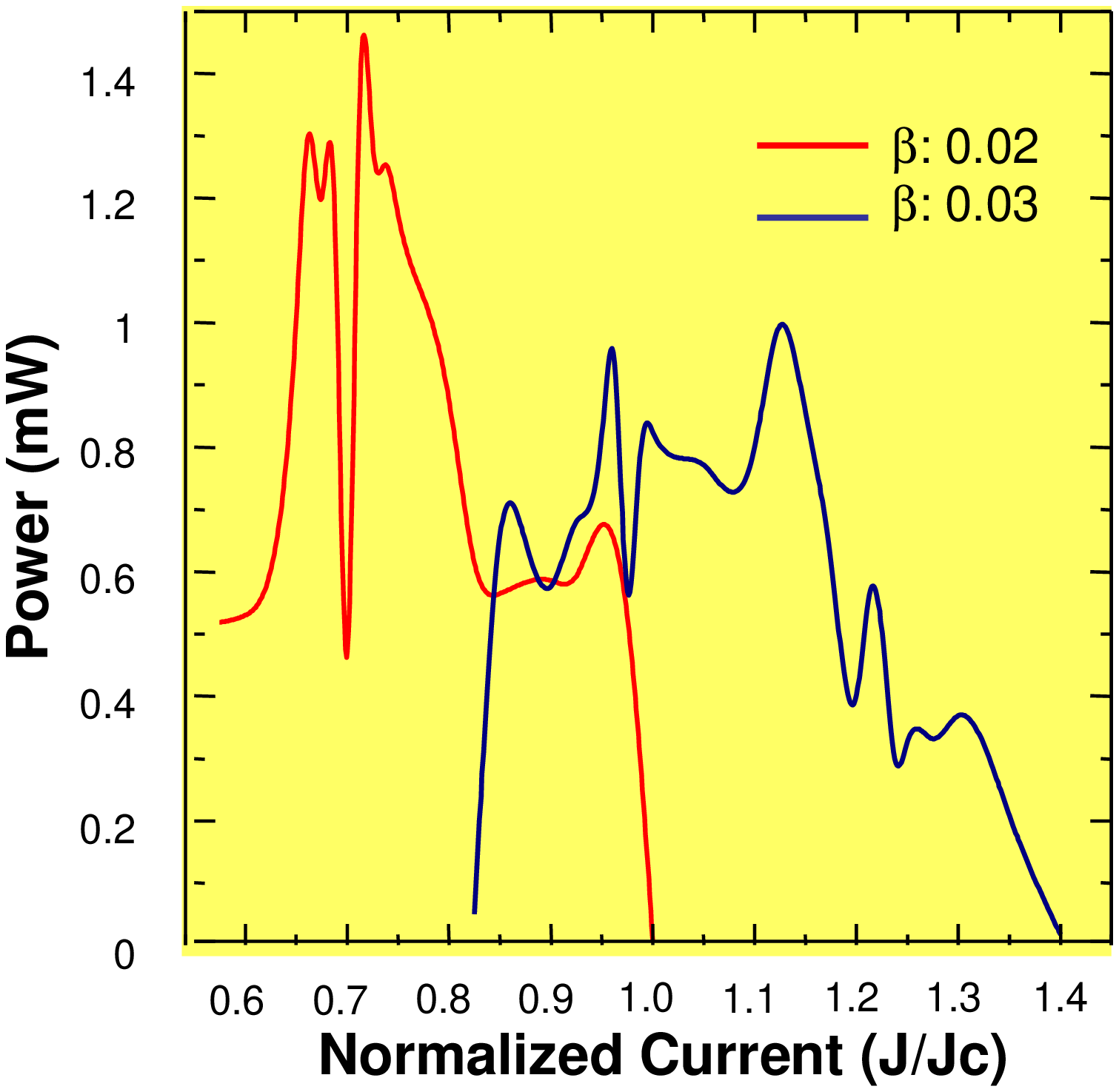}
\caption{The emission power (mW) of the electromagnetic wave as a function of the external current.}
\end{figure}

Figure 5 shows the emission power (the Poynting vector) measured at the location 4$\mu$m from the interface as a function of the external current. The length of the sample along the y-axis is 500$\mu$m. The emission occurs in some limited current region as seen in Fig5. When the current is smaller than a critical current, the emission power vanishes. The reason is that nodes in the Josephson plasma appear along the z-axis, indicating inclusion of the longitudinal component, and thus the emission power vanishes as mentioned before. When the current is larger than the upper critical value, the fluxons take various dynamical orders\cite{Koshelev1}. In this state, the emission vanishes since the rigid fluxon lattice flow prohibits occurrence of the standing wave. The sharp dip of power appearing at $J/J_{c}$=0.7 for $\beta$=0.02 comes from the following reason. The fluxons flow with a small amplitude vibration. When the Josephson frequency resonates with the fluxon vibration frequency and this resonance stabilizes the Josephson plasma, it depresses the conversion rate from the Josephson plasma to the electromagnetic wave. When $\beta$ decreases, for example $\beta$=0.01, the oscillations of electric and magnetic fields lose the coherence and the emission power vanishes.  When $\beta$ increases, the dumping becomes larger and the emission power decreases.

In the present paper we performed the simulation of terahertz wave emission with the parameters corresponding to Bi$_{2}$Sr$_{2}$CaCu$_{2}$O$_{8+\delta}$, and obtained the following results: The sample works as a cavity, and the input energy is stored in a form of standing wave of the Josephson plasma. A part of the energy is emitted as terahertz waves. The emission of electromagnetic waves is in the form of continuous coherent terahertz waves with mW power in the present model. The frequency is tunable by changing the applied current.

\begin{acknowledgments}

We thank Dr. T. Sato to permit us for using the Earth Simulator and Dr. Y. Kanada for the benefit of our use of SR8000. We thank Dr. T. Yamashita, Dr. T. Hatano, Dr. Y. Takano, Dr. K. Hirata, Dr. K. Kadowaki, Dr. H. Wang, Dr. S. Kim, Dr. T. Tachiki, Dr. T. Koyama, Dr. M. Machida, H. Matsumoto, and G. Prabhakar for valuable discussions.
\end{acknowledgments}


\end{document}